# Phase-controlled dual-comb coherent anti-Stokes Raman spectroscopic imaging

Kun Chen[1†], Tao Wu[1†], Tao Chen[1], Tian Zhou[1], Haoyun Wei[1,*] & Yan Li[1,*]

[1]Key Lab of Precision Measurement Technology & Instrument, Department of Precision Instrument, Tsinghua University,

Beijing 100084, China

† These authors contributed equally to this work.

*e-mail: liyan@mail.tsinghua.edn.cn; luckiwei@mail.tsinghua.edu.cn

## Abstract

Coherent Raman microscopy provide label-free imaging by interrogating the intrinsic vibration of biomolecules. Nevertheless, trade-off between high chemical-specificity and high imaging-speed currently exists in transition from spectroscopy to spectroscopic imaging when capturing dynamics in complex living systems. Here, we present a novel concept in dual-comb scheme to substantially beat this trade-off and facilitate high-resolution broadband coherent anti-Stokes Raman spectroscopic imaging based on down-converted, automatically varying delay-time in spectral focusing excitation. A rapid measurement of vibrational microspectroscopy on sub-microsecond scale over a spectral span ~700 cm$^{-1}$ with solid signal-to-noise ratio provides access to well-resolved molecular signatures within the fingerprint region. We demonstrate this high-performance spectroscopic imaging of spatially inhomogeneous distributions of chemical substances as well as the carotenoids accumulation in plant cells. This technique offers an unprecedented route for broadband CARS imaging with hundreds of kHz frame rate using available 1-GHz oscillators.

## Introduction

Coherent Raman spectroscopic imaging has emerged as a high-specific optical tool interrogating vibrational transitions in molecules to enable label-free biomedical microscopy. Successful techniques for coherent Raman microscopy include coherent anti-Stoke Raman scattering (CARS) microscopy [1] and stimulated Raman scattering (SRS) microscopy [2]. Indeed, video-rate imaging of single Raman bands have been achieved with narrow spectral bandwidth by using ultrafast lasers [3-4]. However, real-time identifying and mapping for multiple or possibly unknown molecules requires broad spectral breadth, desired resolution and high-speed acquisition, and has been a long-term pursuit in the field [5-23]. Multiplex or broadband coherent Raman spectroscopy, in general, contains considerably more information and is consequently more useful for qualitative discrimination of many overlapping peaks [5-6]. In particular, multiple peaks within the weakly scattering Raman fingerprint region (<1,800 cm$^{-1}$) are crucial to discriminate subtly different states of cells and tissues [5]. There has been a great effort in developing high-speed multiplex spectroscopic imaging techniques. Hyperspectral SRS and CARS microscopy are now within reach using frequency-swept narrowband laser beams for acquiring images at a series of Raman shifts, but suffer from limitations in laser tuning speed and are not applicable to highly dynamic objects [7-11]. Broadband CARS microscopy using parallel-detection can simultaneously excite multiple Raman bands and reach a speed of 3.5 ms for one pixel [5]. Parallel-detection based multiplex SRS technique has also been demonstrated at the speed of microsecond scale [12]. In contrast to the parallel-detection technique, modulation multiplexing CARS and SRS techniques use a point detector, thus enabling high sensitivity and speed [13-18]. It generally involves a single femtosecond laser and a mechanical motor for stepping the delay, and the latest techniques enabled by a high-speed resonant-scanner line has achieved tens of microseconds acquisition time [19, 20, 15, 17]. However, mechanical scanner fundamentally limits the measurement time, spectral resolution and system stability. Femtosecond optical frequency combs offer an intriguing solution to the rapid-scanning problem. From the early work by Newbury and his colleagues [21], combs have been incorporated into broadband molecular linear absorption spectroscopy, constituting a virtual scanning interferometer. In particular, by exploiting their potential for nonlinear spectroscopy, combs have been harnessed for CARS imaging [22]. This motionless frequency-comb-based technique enables more than 1,000-fold shorter acquisition time, followed by Fourier transformation to obtain spectral information. All spectral elements were acquired within as short as 12 μs with 100-MHz oscillators, while at a high resolution. However, actual refresh rate is up to tens and hundreds of milliseconds, limiting successive spectral acquisitions and imaging. Larger repetition frequency difference can lead to increase in refresh rate, but dramatically decrease signal-to-noise ratio (SNR). One solution would be to use combs with larger spacing lines which has been demonstrated very recently with 1-GHz oscillators [23]. The refresh rate is improved in proportion to the repletion rate of laser and the measurement time may get down to 5 μs, while the spectrum has a SNR about one order of magnitude worse than that measured with the 100-MHz system. Further improvement of the imaging rate for high dynamic objects needs 10-GHz oscillators, putting towards a high request for high-power laser and high-speed detector. Though considerable efforts have been devoted to the establishment and development of nonlinear dual-comb Raman spectroscopy, unveiling the full potential of comb-based coherent Raman microscopy still calls for advanced strategies.

In this article, we report an ultra-rapid broadband CARS imaging system by using two linearly chirped 100-MHz combs with two orders of magnitude improvement in both refresh time and measurement time over previously experiments. This phase-controlled excitation offers a flexible route for optimizing the refresh rate and spectral resolution, while circumventing the trade-off between high SNR and improved imaging rate. More importantly, it reveals the prospect up-front of the proposed approach that one can substantially achieve up to hundreds of kHz imaging rate with available higher repetition frequency oscillators, which might pave the way towards chemical imaging of highly dynamic objects.

## Results

**Measurement concept.** CARS is a vibrationally enhanced four-wave mixing process which can be incorporated within a single broadband pulse by controlling the spectral phase to determine the population of vibrational levels and restore high spectral resolution. One well-established method is to apply sinusoidal spectral phase functions to create a pulse train, but spectral resolution was obtained only after measuring with pulse trains of different periods, incompatible with a fast acquisition. [13, 24-26]. Control of the excitation process is also achieved by applying a quadratic spectral phase modulation, called linear chirp, to the broadband input pulse and then splitting it into two beams [27-28]. Scanning the time delay between the two chirped pulses can selectively populate different vibrational energy levels. This is known as the spectral focusing approach [29], and there have been great efforts establishing and developing of this phase-controlled technique in the past decade [14, 17-18, 30-36].

In our experiment, in the time-frequency plane (Fig. 1a), transform-limited pump and Stokes pulses are linearly stretched in the time domain with a common chirp parameter $\alpha$, by applying the equal quadratic phase to each pulses. The composite light field of the pump and Stokes pulses then creates multiple pairs of constant, narrowed, instantaneous frequency differences (IFD). In the phase-controlled dual-comb CARS, we harness two femtosecond lasers with repetition rate $f_r$ and $f_r$-$\delta f_r$ to excite a sample (Fig. 1b). At each excitation, the pulses from each comb source overlap on the sample at a linearly increasing delay time, automatically induced by the small difference in repetition frequencies $\delta f_r$. A full 'scan' of the pump pulse across the Stokes pulse is accomplished every 1/ $\delta f_r$ in real measurement time and every 1/ $f_r$ in delay time with a down-conversion factor of $\delta f_r$ / $f_r$. With this asynchronous optical sampling, delay time and IFD are connected in a well-defined way by the chirp parameter $\alpha$. Thus, automatic and ultra-rapid scanning of the delay-time permits selective Raman mode excitation and sequentially capture all the spectral elements with ultra-short measurement time as shown in Fig. 1c. Spectral resolution is calculated by multiplying the down-converted delay-time interval of $\delta f_r$ / $f_r^2$ and chirp parameter $\alpha$. Refresh rate ($\delta f_r$) is therefore proportional to the square of repetition frequency ($f_r$) when a desired and definite resolution is chosen. More importantly, the signal intensity for each excitation will remain consistent by concentrating most of the optical power into a single Raman mode no matter what the repetition frequency difference is. Thus, one can achieve high refresh rate without significant loss SNR and further improve the rate up to hundreds of kHz with higher repetition frequency combs.

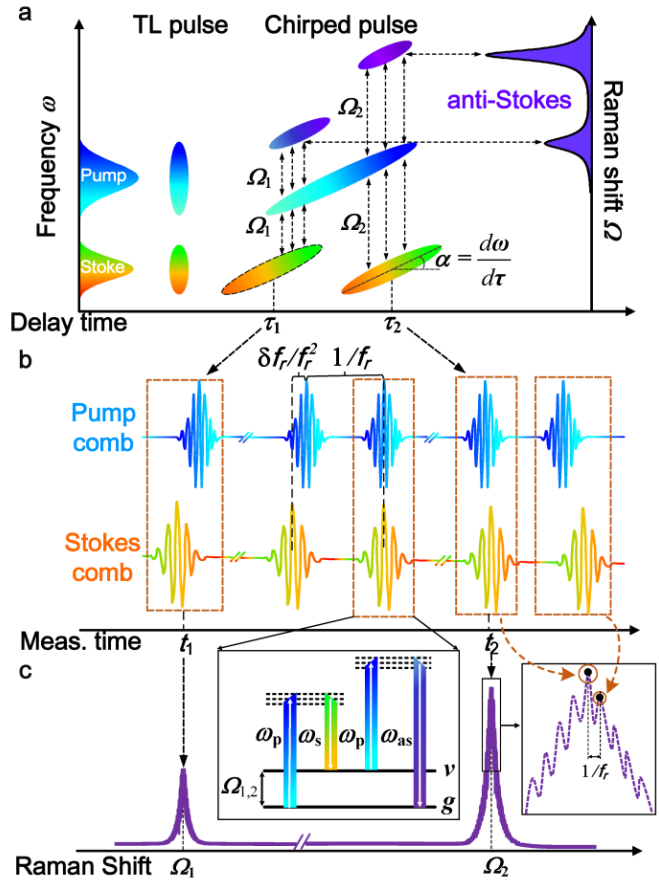

**Figure 1 | Measurement concept of phase-controlled dual-comb CARS. a,** Phase-controlled excitation enables high spectral resolution imaging and fast spectral tuning. Ultrashort transform-limited (TL) pulses can be stretched in time domain by applying quadratic phase. The composite light field of pump and Stokes pulse, which are equally chirped, creates multiple pairs of constant, narrowed, instantaneous frequency difference. By changing the pulse delay between the pump and Stokes pulses can selectively populate a given vibrational Raman level ($\Omega_1$ or $\Omega_2$) at different delay time ($\tau_1$ or $\tau_2$), generating high-frequency anti-Stokes signals. $\alpha$ is chirp parameter. **b,** Ultra-rapid and automatic scanning of delay time enables to probe multiple characteristic vibrational levels in an asynchronous optical sampling. The Stokes generates pulses at a slightly different repetition rate with respect to the pump. Every repetition period ($1/f_r$), the Stokes pulse 'slips' by $\delta f_r / f_r^2$ relative to the pump pulses, and creates a slightly different instantaneous frequency difference. **c,** Consecutive excitation and acquisition by the overlap of pump and Stokes pulses yield a high-speed and high-resolution measurement of broadband CARS spectroscopy. Actual data are shown on the right side, where each discrete point corresponds to a single digitized sample when the pump and Stokes pulses overlap on the sample and only the immediate overlap region is shown. , Sequential $1/f_r$ traces are placed side by side to map multiple characteristic vibrational frequencies (see the inserted larger boxed regions on the right side). Insets showing energy-level scheme of the phased-controlled dual-comb CARS process. $\omega_p$, $\omega_s$ and $\omega_{as}$ are the frequency of pump, Stokes and anti-Stokes, respectively.

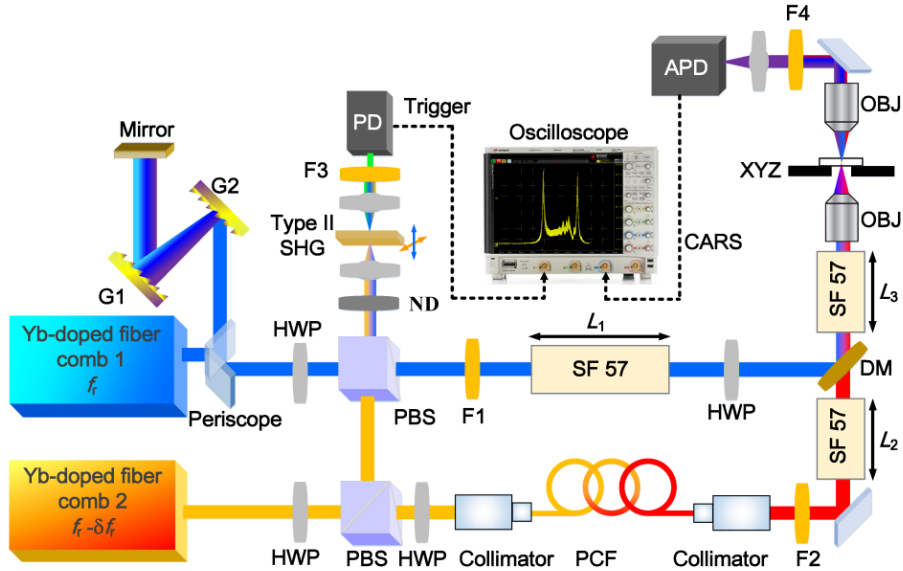

**Figure 2 | Schematic of the experimental set-up.** See Methods for details. HWP, half wave-plate; PBS, polarizing beam splitter; PCF, photonic crystal fiber; F1, F2 long-pass filter; F3, F4, short-pass filter; G1, G2, grating pairs; SF57, SF57 glass rod; DM, dichroic mirror; OBJ, objective lenses; Pol, polarizer; SHG, second harmonic generation; ND, neutral density filter; PD, photo diode; APD, avalanche photo diode; XYZ, piezoelectric stage; $L_1$=30 cm, $L_2$=5 cm, $L_3$=15 cm. Insets is a real screen shot of the dual-comb CARS spectrum acquired by an oscilloscope (DSO-S 254A, KEYSIGHT).

Figure 2 presents a schematic of the phase-controlled dual-comb CARS experimental setup (see Methods). Here, the pump wavelength is centered at 1060 nm while the Stokes comb redshifts to about 1260nm so as to access the chemically rich and biologically relevant fingerprint region (1100-1800 cm$^{-1}$) [35-36]. Second harmonic generation (SHG) servers as a trigger signal to acquire CARS spectra and calibrate vibrational frequency. It does not affect the quality of an individual spectrum but enables the high reproducibility of the wavenumber scale of a sequence of spectra and provides accurate determination of the vibrational levels. High-index glass rods basically stretch a pulse durations to several picoseconds that is comparable to the relevant Raman coherence time.

**Phase-controlled dual-comb CARS spectroscopy.** The time-domain Raman spectra signal is periodic. Every $\delta f_r$, a strong burst contains most of the spectral elements resulting from the varying overlap of the dual-comb pulses. The time-windowed portion of the burst is only determined by the spectral bandwidth of the laser pulses. Sequential $1/f_r$ traces are placed side by side to map the entire CARS spectrum of retinoic acid (RA), where the carrier pulses with 100 MHz repetition rate is modulated by molecular transitions (Fig. 3a). Demodulation is able to extract the CARS spectra only by using a low-pass filter or profile-extraction algorithm. In order to remove the intensity variation at low and high wavenumbers due to decreased pulse overlap, calibration against the known spontaneous Raman signals can be implemented if needed [18]. Additionally, the CARS signal generally presents slightly dissipative spectral lineshape, which is ascribed to the weak nonresonant backgrounds and well-understood in the CARS community. Because the dual-comb CARS is a spectroscopic technique, vibrational phase, which excludes the nonresonant contribution, can be extracted mathematically from the CARS signal intensity [5, 37-38]. Fig. 3c shows the intensity-calibrated and baseline-corrected dual-comb CARS spectrum, in which the spectral positions and FWHMs of different Raman resonance match very well with the spontaneous spectrum (Fig. 3b).

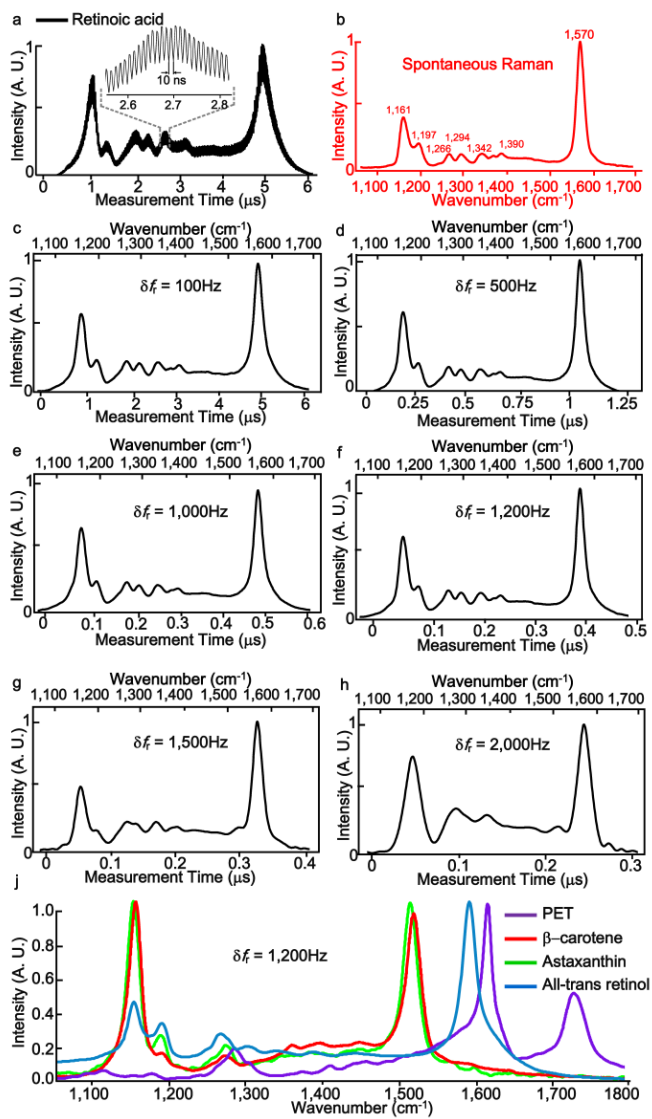

**Figure 3 | High-resolution phase-controlled dual-comb CARS of different chemicals within ultra-short measurement time. a,** Unaveraged original time-domain CARS signal of RA, where the carrier pulses with 100 MHz repetition rate is modulated by the vibrational transitions. **b,** Spontaneous spectrum of RA. **c.** The intensity-calibrated and baseline-corrected dual-comb unaveraged CARS spectrum of RA by demodulating the time-domain traces in a ($\delta f_r$ =100 Hz; measurement time 6 μs; estimated resolution 3.5 cm$^{-1}$); **d.** Dual-comb unaveraged CARS spectrum of RA ($\delta f_r$ =500 Hz; measurement time 1.2 μs; estimated resolution 5 cm$^{-1}$); **e.** Dual-comb unaveraged CARS spectrum of RA ($\delta f_r$ =1,000 Hz; measurement time 0.6 μs; estimated resolution 10 cm$^{-1}$); **f.** Dual-comb unaveraged

CARS spectrum of RA ($\delta f_r$ =1,200 Hz; measurement time 0.5 μs; estimated resolution 12 cm$^{-1}$); **g.** Dual-comb unaveraged CARS spectrum of RA ($\delta f_r$ =1,500 Hz; measurement time 0.4 μs; estimated resolution 15 cm$^{-1}$); **h.** Dual-comb unaveraged CARS spectrum of RA ($\delta f_r$ =2,000 Hz; measurement time 0.3 μs; estimated resolution 20 cm$^{-1}$); **i.** Dual-comb unaveraged CARS spectra of PET film, β-carotene, astaxanthin and all-trans retinol spanning Raman shifts from 1,100 cm$^{-1}$ to 1,800 cm$^{-1}$ ($\delta f_r$ =1200 Hz; measurement time 0.57 μs; estimated resolution 12 cm$^{-1}$); The energy per pulse of pump and Stokes pulses after the focusing microscope objective are 0.45 nJ and 0.05nJ.

We illustrate acquisition times with five spectra recorded with $\delta f_r$ =100 Hz (Fig. 3c), 500 Hz (Fig. 3d), 1,000 Hz (Fig. 3e), 1,200 Hz (Fig. 3f), 1,500 Hz (Fig. 3g) and 2,000 Hz (Fig. 3h) for RA. The spectra involve no averaging and were measured in 6 μs, 1.2 μs, 0.6 μs, 0.5 μs, 0.4 μs and 0.3 μs from Fig. 3c to h. It is clear that increase of the repetition frequency difference will decrease measurement time, but also decrease the spectral resolution. The resolution can be estimated as 3.5 cm$^{-1}$ with $\delta f_r$ =100 Hz and as 20 cm$^{-1}$ with $\delta f_r$ =2,000 Hz. Fig. 3f demonstrates that one can still faithfully retrieve all the Raman signatures of RA for a refresh rate of 1.2 kHz with resolution of 12 cm$^{-1}$, showing almost fairly consistent spectral contours and intensities compared with the spontaneous spectrum. Both refresh time and measurement time achieve one to two orders of magnitude improvement in than previously possible. More importantly, the SNR of the spectra, recorded under different refresh rate, remains consistent. It culminates at about 400 for the most intense blended line of RA at 1570 cm$^{-1}$ in Fig. 3c-h. Because SNR is immune to the increase of refresh rate, it is very advantageous to optimize the refresh rate and resolution of CARS spectra, in general, and to optimize the imaging speed and chemical specificity of CARS microscopy, in particular.

We further performed dual-comb CARS measurements on polyethylene glycol terephthalate (PET) film, β-carotene, astaxanthin and all-trans retinol spanning Raman shifts from 1,100 cm$^{-1}$ to 1,800 cm$^{-1}$ within the chemically rich fingerprint region. One can identify significant spectral complexity and specificity of different samples with about 0.57 μs measurement time at a refresh rate of 1.2 kHz. This system provides a level of signal clarity throughout the fingerprint region previously only available through spontaneous Raman, but at speeds aligned with those of coherent Raman techniques. Multiple characteristic vibrational Raman signatures significantly enhance the ability to discriminate between different substances, and benefit in-line quantitative analysis for composition information.

**Phase-controlled dual-comb CARS chemical imaging.** To date, vibrational spectroscopic imaging for biology and medicine has relied on limited spectral information, primarily in the strong CH/OH stretch region of the Raman spectrum (∼2,800–3,500 cm$^{-1}$), without adequate chemical specificity and sensitivity to distinguish a variety of substances [3,6, 8, 18]. To demonstrate the imaging capability of this approach to identify multiple molecular fingerprint signatures, we imaged a mixture of RA crystal and β-carotene. RA contributes in formation of photopigment, and β-carotene is a precursor in RA synthesis, they function to improve visual performance. Originated from the similar structure, RA and β-carotene obtained alike Raman signatures, which requires wide spectral information for discrimination. The nonlinearity of CARS permits intrinsic optical sectioning capability. We dissected the three-dimensional distribution of RA and β-carotene in a 100 μm × 100 μm× 22 μm space, with 1 μm × 1 μm× 1 μm voxel. In the reconstructed pseudocolour image, RA was highlighted in yellow (1570 cm$^{-1}$, C = C stretch) and β-carotene was in red (1521 cm$^{-1}$, C = C stretch). The local distribution of RA and β-carotene can be inspected from arbitrary section, as shown in Fig. 4b. Two reconstructed axial planes are also shown. For each pixel, we measure all the spectral elements from 1,100 to 1700 cm$^{-1}$ within 0.5 μs. The total measurement time of 8.3 s corresponds to an acquisition rate of 1200 pixels s$^{-1}$. Spectra of RA and β-carotene recovered from single pixels (marked in Figure 4b), A, B C, and D were shown in Figure 4c, which provides rich chemical information to clearly discriminate the two substances at a resolution of ~12cm$^{-1}$. It more clearly shows the shape, size and orientation of the individual constituents in the reconstructed three-dimensional image of Fig. 4a. Though optical sectioning capability has been demonstrated in previous narrowband CARS and SRS researches [1, 4, 8], three-dimensional microspectroscopy with broadband CARS has still been non-trivial due to their long acquisition times. Our dual-comb approach allows fast three-dimensional microspectroscopy and microscopy.

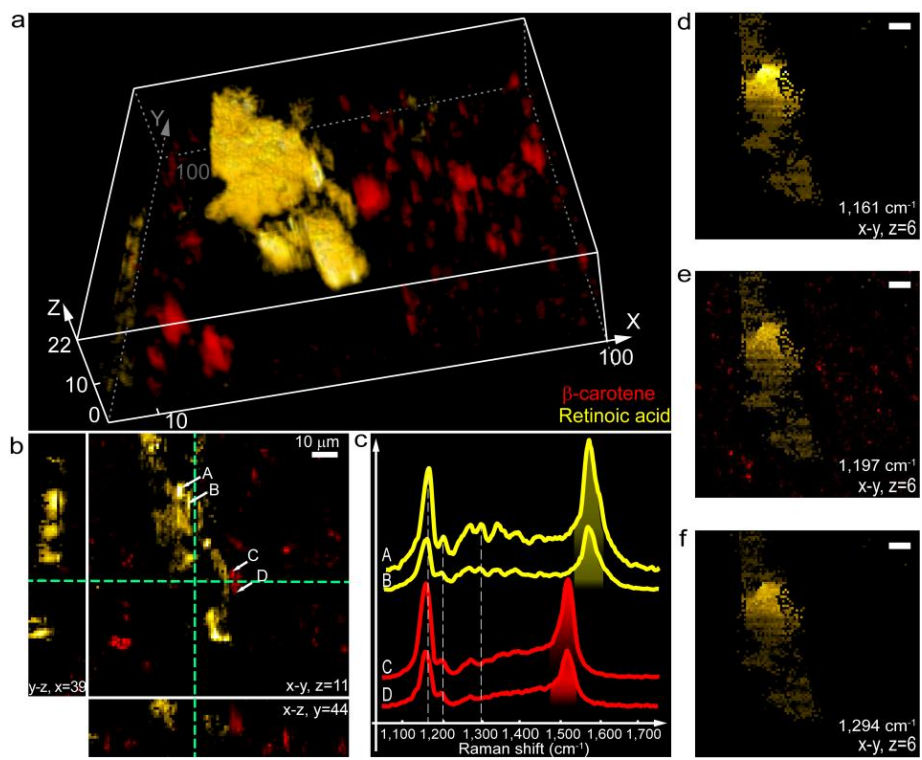

**Figure 4 | Three-dimensional dual-comb CARS imaging of the mixture of RA and β-carotene. a,** Three-dimensional reconstruction of distribution of RA and carotene from twenty-two z-stack images, with the RA highlighted in yellow (1570 cm$^{-1}$) and β-carotene in red (1521 cm$^{-1}$) (100 pixels × 100 pixels, 0.5 μs measurement time, 1.2 kHz refresh rate). **b,** Pseudocolour image taken from a single plane of

z-stack image collection in a. **c,** Single-pixel spectra taken from RA and β-carotene, marked as A, B, C and D in b. **d-f,** Additional spectral channels that provide chemical contrast: 1,161 cm$^{-1}$ but not 1520 cm$^{-1}$ (**d**); 1,197 cm$^{-1}$ (**e**) and 1,294 cm$^{-1}$ (**f**). Each channel was marked by dash line in **c**. Scale bars, 10 μm.

The broad Raman band recorded by phase-controlled dual-comb CARS showed significant spectral complexity in the sample, as illustrated by the single-pixel spectra in Fig. 4c. Using isolated peaks, one could create dozens of unique images based on different vibrational susceptibilities, such as those shown in Fig. 4d–f: 1,161 cm$^{-1}$ ($\upsilon$(C–C)), 1,197 cm$^{-1}$ ($\upsilon$(C–C)) and 1,294 cm$^{-1}$ ($\rho$(CCH)), respectively. Additionally, a multivariate analysis of contributions from several peaks—their locations, intensities and shapes—presents significant avenues of chemical contrast. For example, Fig. 4d highlights RA by segmenting the chemical species that have vibrations at 1,161 cm$^{-1}$ but lack vibrations at 1,520 cm$^{-1}$, which isolates RA from β-carotene. Additionally, with this spectroscopic technique, the SNR and contrast of the CARS images could be further improved with a more intricate intensity-extracting algorithm, benefiting from its multiplex property.

Beyond the mixture of pure RA and β-carotene, we then applied the system to native plant tissue of bell pepper, to specifically image the intracellular carotenoid, which is non-trivial. Carotenoids, which include carotenes and their oxygen-containing derivatives - xanthophylls, play many important biological functions in plants. Bell pepper primarily contains three naturally occurring carotenoids, lutein, β-carotene and zeaxanthin, giving two intense bands near 1525 cm$^{-1}$ and 1157 cm$^{-1}$, which can be attributed to stretching modes of conjugated C=C and C-C bonds in the central chain, respectively.

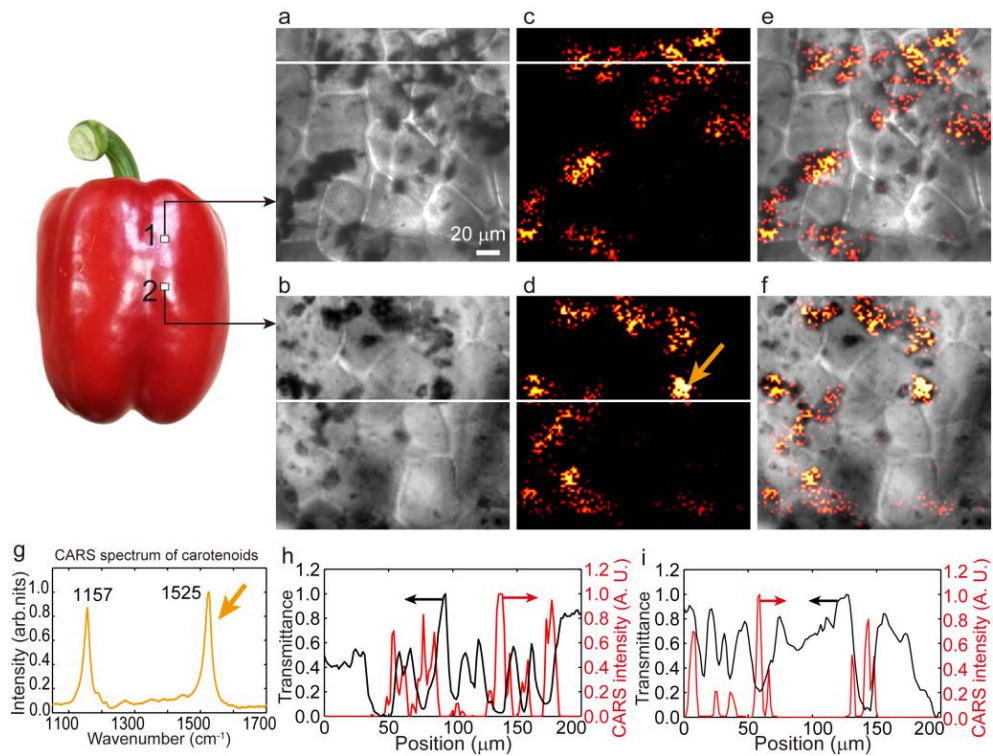

**Figure 5 | Dual-comb CARS imaging of carotenoids accumulation in bell pepper pericarp tissue sections. a-b,** The high-definition transmission microscopy image from the pump laser of the same setup in section 1 and 2 of the bell pepper pericarp tissue, area 200 μm ×

200 μm. **c-d,** CARS microscopy images of carotenoids accumulation with it highlighted in red using 1510-1530 cm$^{-1}$ for contrast in the same field as a and b, respectively. (100 pixels × 100 pixels, 0.5 μs measurement time, 1.2 kHz refresh rate) **e-f,** The merged image of a and c, and, b and d, respectively. **g,** Single-pixel spectra taken from carotenoids, marked with arrows in d. **h-i,** Corresponding cross-sectional profiles of the images along the indicated white line in a and c, b and d: (**h**) corresponds to a (black line) and c (red line); (**i**) corresponds to b (black line) and d (red line).

Two bell pepper pericarp tissue sections, hand-dissected from areas between inner placental veins (secondary phloem regions), were first observed by the transmission images of the pump beam (Fig. 5a-b). Figure 5c and d show the CARS microscopy images of carotenoids accumulation highlighted in red using the integrated intensity of 1520-1530 cm$^{-1}$ for contrast, which was collected with 0.5 μs measurement time over a 200 μm × 200 μm area (100 pixels × 100 pixels). For each pixel, we measure the spectrum from 1,100 to 1700 cm$^{-1}$. The two images clearly demonstrate there is a diversity in fruit pericarp chromoplast architecture and associated differences in carotenoid content. The merged image of the transmission and CARS of pericarp (Fig. 5e and f) reveal that the carotenoids in plastids of plants have low transmittance and high absorbance in near infrared region (1,060 nm) where the dark black regions of transmission image significantly overlap the carotenoids accumulation areas. The corresponding cross-sectional profiles of the images further demonstrate considerably match between the low transmittance and high CARS signal regions. This multi-modal image allows to simultaneously characterize the sample morphology with very high detail and the molecular composition with functional specificity. Additionally, one can use a cross correlation algorithm to retrieve the weak signals of carotenoids from low-accumulation regions, taking advantage of its broadband CARS spectrum. We demonstrate a significant image contrast improvement for carotenoids distributions in Fig. S5, which can improve the interpretation of local chemical structures and distribution. Because a number of carotenoids act as vitamin A precursors for human beings, the accumulation of high levels of carotenoids is of great interest due to their antioxidant properties and potential impact on human health [38, 39]. Our phase-controlled dual-comb CARS thus provides a high-specificity microscopy techniques for the monitoring of the subcellular distribution and accumulation of provitamin A carotenoids in plant cells.

## Discussion

In this work, we demonstrate an ultra-rapid broadband CARS imaging system by using two linearly chirped 100-MHz combs, achieving up to two orders of magnitude improvement in both refresh time and measurement time. The realization of such achievements is obtained through rapid-scanning of down-converted delay-time in conjunction with ultrafast acquisition of asynchronous optical geometry. This phase-controlled excitation circumvents the trade-off between high SNR and improved refresh rate by concentrating most of the tailored optical driver fields into a single Raman mode. One can therefore achieve high refresh rate without significant loss SNR and further improve the rate with higher repetition frequency combs. Multiple vibrational frequencies within the chemically rich fingerprint region can be efficiently probed in timess as short as hundreds of nanoseconds. High-specificity CARS microspectroscopy and microscopy are implemented using a single photodetector to demonstrate the superior performance of the proposed approach.

Since the CARS spectral resolution is mainly set by the comb repetition rates, rather than by mechanical motion as in a canonical spectral focusing geometry, phase-controlled approach is a more promising alternative to alter the chemical selectivity of different vibrational bands in a relatively compact, easily aligned system. All-fiber laser source is also advantageous over the free space lasers in terms of its simplicity, flexibility and lower cost. Additionally, our method allows intrinsically feasible tunability of Stokes pulse based on the soliton self-frequency shift, and we can obtain fingerprint vibrational signatures over a large spectral range (800-2200 $cm^{-1}$), while the long wavelengths of both light sources are very attractive for deep tissue imaging. High-power soliton-induced supercontinuum generation can benefit further applications in biology and materials [40]. Advanced phase-controlled scheme have recently emerged, and we believe that their combination with our method will deliver techniques with improved performance and utility [34, 41]. In addition to the aforementioned aspects, microresonator dual-comb system may offer a route towards real clinical diagnosis [42].

Though our method achieve measurement time on sub-microsecond scale and refresh rate of kHz-level with 100-MHz oscillators, Raman spectroscopic imaging of highly dynamic organelles in live cells or fast flowing objects still calls for at least one order of magnitude improvement in refresh rate. In the recent dual-comb scheme by using 1-GHz frequency oscillators, Mohler and his co-workers reported a refresh rate of 2 kHz, and they also believe that several tens of kHz refresh rate will be within reach with 10-GHz combs in the future [23]. Repetition frequencies above 1-GHz provides a straightforward way to speed up the imaging process, but makes the nonlinear excitation more challenging and necessitates a high-speed and high-sensitivity detector, since the CARS signal scales with the cube of the peak power of the laser pulses. With down-converted delay-time in spectral concentrating excitation, our method offers an intriguing prospect up-front that one can expect to achieve broadband Raman spectra measured on tens of nanoseconds scale at a refresh rate up to hundreds of kHz, while keeping solid resolution and SNR, with the now availability of commercial 1-GHz repetition frequency oscillators. Thus, phase-controlled dual-comb CARS might now have the most advanced potential to evolve into a powerful tool for label-free chemical imaging of highly dynamic organelles, cellular states and cellular processes, in live cells.

## Acknowledgments

This work is funded by the National Natural Science Foundation of China (Grant No. 51575311) and the State Key Lab of Precision Measurement

Technology & Instrument, Tsinghua University. We acknowledge support from Andrew A. Pascal at Université Paris Saclay, Gif-sur-Yvette, France for fruitful discussion about the plant samples.